\documentclass[aps,prb,reprint,twocolumn,amsmath,longbibliography]{revtex4-2}
\usepackage{bbm}
\usepackage{mathrsfs}
\usepackage{amsmath}
\usepackage{amsfonts}
\usepackage[colorlinks=true, 
            citecolor=blue, 
            anchorcolor=blue, 
            linkcolor=blue, 
            urlcolor=blue]{hyperref}
\usepackage{graphicx,epstopdf}
\usepackage{subfigure}
\usepackage{epsfig}
\usepackage{dcolumn}
\usepackage{bm}
\usepackage{color}
\usepackage{natbib}
\usepackage{amssymb}
\usepackage{xcolor}
\usepackage{braket}
\begin{document}

\title{Electrically tunable high-Chern-number quasiflat bands in twisted antiferromagnetic topological insulators}

\author{Huaiqiang Wang$^{1}$, Yiliang Fan$^{2}$ and Haijun Zhang$^{2,3,\ast}$}

\affiliation{
 $^1$ Center for Quantum Transport and Thermal Energy Science, School of Physics and Technology, Nanjing Normal University, Nanjing 210023, China\\
 $^2$ National Laboratory of Solid State Microstructures, School of Physics, Nanjing University, Nanjing 210093, China\\
 $^3$ Collaborative Innovation Center of Advanced Microstructures, Nanjing University, Nanjing 210093, China\\
}

\begin{abstract}
Isolated flat bands with significantly quenched kinetic energy of electrons could give rise to exotic strongly correlated states from electron-electron interactions. More intriguingly, the interplay between topology and flat bands can further lead to richer physical phenomena, which have attracted much interest. Here, taking advantage of the recently proposed intertwined Dirac states induced from the anisotropic coupling between the top and bottom surface states of an antiferromagnetic topological insulator thin film, we show the emergence of a high-Chern-number (quasi)flat-band state through moir\'e engineering of the surface states. Remarkably, the flat bands are isolated from other bands and located near the Fermi level. Furthermore, topological phase transitions between trivial and nontrivial flat-band states can be driven by tuning the out-of-plane electric field. Our work not only proposes a new scheme to realize high-Chern-number flat-band states, but also highlights the versatility of the intertwined Dirac-cone states.
\end{abstract}

\email{zhanghj@nju.edu.cn}

\maketitle

\section{Introduction} 
Recently, the seminal experimental findings of unconventional superconductivity and correlated insulator behavior in twisted bilayer graphene (TBG)~\cite{cao2018correlated, cao2018unconventional} have ignited a surge of research interest in condensed-matter systems hosting flat-band electronic structures. In these flat bands, the kinetic energy of electrons is significantly quenched, and electron-electron interactions become dominant, which could give birth to exotic strongly-correlated states, such as the fractional quantum Hall effect (FQHE)~\cite{Stormer1999Fractional}, fractional Chern insulators, and fractional topological insulators (TIs)~\cite{Levin2009Fractional, Maciejko2010Fractional, Qi2011Generic, Sun2011Nearly, Tang2011High, Neupert2011Fractional, Stern2016Fractional, spanton2018observation}. Interestingly, it has been shown that flatbands  possessing a high Chern number ($C>1$) could generate many new states \cite{Yang2012Topological, Sterdyniak2013Series, Barkeshli2012Topological, Trescher2012Flat} beyond the Landau-level-like case with $C=1$ in the FQHE. This renders realistic platforms possessing isolated high-Chern-number flat bands very attractive and highly desired~\cite{bao2024isolated}.

As for the realization of flat-band states, moir\'e engineering of two-dimensional van der Waals heterostructures, e.g., TBG~\cite{bistritzer2011moire, andrei2020graphene} and twisted transition metal dichalcogenide multilayers~\cite{tang2020simulation, regan2020mott, shimazaki2020strongly, wang2020correlated}, has proved to be a versatile tool, where the twist angle plays a very crucial role. Moreover, moir\'e engineering has also been applied to other Dirac materials, in particular the Dirac-cone surface state (DSS) of three-dimensional TIs~\cite{Cano2021Moire,Dunbrack2022Magic,Wang2021Moire}. However, since the DSS in TIs is anomalous and protected by time-reversal symmetry, it remains gapless and cannot be disconnected from other bands under the moir\'e superlattice potential. To circumvent this issue, a natural way is to gap the DSS by magnetism, from which isolated moir\'e flatbands could be obtained~\cite{Liu2022Magnetic, Chaudhary2022Twisted}. To this end, the recently discovered intrinsic magnetic TI MnBi$_2$Te$_4$ and its family materials~\cite{Gong2019cpl, Otrokov2019nature, Zhang2019mbt, Li2019sa, Chen2019intrinsic, chen2019prx-ARPES,li2019prx-ARPES,hao2019prx-ARPES,Otrokov2019prl, Sun2019rational, Vidal2019topological,Ge2020high, Hu2020van,Wang2020dynamical, Fu2020exchange,Lian2020prl,Sass2020prl,Liu2020robust,Gu2021spectral, Li2021prl,gao2021layer, liu2021magnetic, Zhu2023Floquet,Bai2023Quantized,Wang2023nature,Gao2023nonlinear,Jo2020,You2021,Du2020} could serve as promising platforms. Intriguingly, when the TI film becomes very thin, the coupling between the top and bottom DSSs cannot be neglected and usually turns out to be critical in the low-energy physics~\cite{Shan2010,Lu2010Massive,Wang2023Dirac,wang2023three}. It is worth mentioning that a recent work~\cite{fan2023intertwined} by some of the authors proposes that anisotropic couplings between the DSSs of an antiferromagnetic (AFM) TI thin film could give rise to emergent new Dirac cones, dubbed \emph{intertwined Dirac-cone states}, away from the $\Gamma$ point. It is thus straightforward to expect that the interplay between magnetism, DSS coupling, and moir\'e engineering can lead to rich new phenomena.

In this work, based on the effective model analysis, we show the emergence of high-Chern-number ($C=n$) flat-band states through moir\'e engineering of the two DSSs of an AFM TI film respecting the $n$-fold ($n=2,3,4,6$) rotational symmetry in the presence of an out-of-plane electric field. The intertwined Dirac-cone states induced from anisotropic couplings are found to play a significant role in this process. Furthermore, the flat bands are electrically tunable, and taking the $n=3$ case as an example, we explicitly demonstrate the topological phase transition between trivial ($C=0$) and nontrivial ($C=3$)  flat-band states. Our work not only proposes a new route to obtain high-Chern-number flat bands, but also highlights potential applications of the intertwined Dirac-cone states.

\begin{figure}[htbp]
    \includegraphics[width=3.2in]{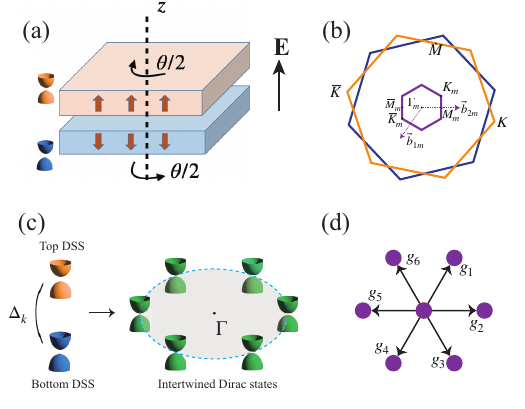}
    \caption{(a) Schematic of the twisted antiferromagnetic topological insulator thin film under an out-of-plane electric field, where a clockwise (counterclockwise) rotation with angle $\theta/2$ is implemented for the top (bottom) surface layer. (b) Two-dimensional Brillouin zones (BZs) of the top (orange lines) and bottom (blue lines) surface layers after the twist, and the moir\'e reciprocal vectors (purple dashed lines) and corresponding moir\'e BZ (purple solid lines). (c) Illustration of the emergent intertwined Dirac-cone states from the coupling between the top and bottom Dirac surface states (DSSs). (d) The first-shell approximation taken in the moir\'e reciprocal lattices for calculating the interlayer moir\'e hopping, where the $g_i$'s are the six smallest moir\'e reciprocal vectors.}
    \label{fig1}
\end{figure}

\section{Model Description}
We consider a thin film of AFM TI with opposite out-of-plane ($z$-axis) magnetic moments on its top and bottom surfaces, e.g., an even-layer MnBi$_2$Te$_4$ film with $A$-type AFM order, as schematically shown in Fig. 1(a). Apart from the combined $\mathcal{PT}$ symmetry from inversion $\mathcal{P}$ and time-reversal operation $\mathcal{T}$, the AFM TI film is assumed to preserve an additional $n$-fold ($n=2,3,4,6$) crystalline rotation symmetry $(C_{nz})$ along the $z$-direction and a combined symmetry $\mathcal{M}_x\mathcal{T}$ from mirror ($\mathcal{M}_x$) and time-reversal ($\mathcal{T}$) operations. When an electric field $E$ is applied along the out-of-plane direction, the precedent $\mathcal{PT}$ symmetry of the AFM TI film is broken. Further, we implement a clockwise (counterclockwise) rotation of angle $\theta/2$ for the top (bottom) surface layer, leading to a relative twisting angle of $\theta$ between them.

To lay a foundation for later discussion, we start from the untwisted case with $\theta=0$, where the low-energy physics of the AFM TI film can be captured by the top and bottom DSSs and the coupling between them. In the ordered basis of $|t,\uparrow\rangle$, $|t,\downarrow\rangle$, $|b,\uparrow\rangle$, $|b,\downarrow\rangle$, where $t(b)$ represents the top (bottom) DSS, the Hamiltonian can be described as
\begin{equation}
H=\left[\begin{array}{cc}h_t +U\sigma_0& h_{\mbox{\tiny{coup}}} \\ h_{\mbox{\tiny{coup}}} ^{\dagger}& h_b-U\sigma_0\end{array}\right],\\
\end{equation}
with 
\begin{equation}
\begin{split}
h_{t(b)}=&\pm\Big[v(k_x\sigma_y-k_y\sigma_x)+m\sigma_z+\frac{R_w}{2}(k_{+}^{n}+k_{-}^{n})\sigma_z\Big],\\
h_{\mbox{\tiny{coup}}}=&(\Delta-Bk^2)\sigma_0-\frac{R_a}{2}(k_{+}^{n}-k_{-}^{n})\sigma_0.\\
\end{split}
\end{equation}
Here, $k=(k_x^2+k_y^2)^{1/2}$ and $k_{\pm}\equiv k_x\pm ik_y$.  $\sigma_i$'s ($i=x,y,z$) are Pauli matrices acting in the spin subspace, and $\sigma_0$ is a $2\times 2$ identity matrix. $U$ is the effective staggered potential caused by the electric field between the top and bottom DSSs. The first term in $h_{t(b)}$ describes the helical DSS, with $v$ denoting the Fermi velocity. The $m\sigma_z$ term represents the Zeeman coupling between the DSS and its surrounding magnetic moment, where the coupling strength $|m|$ is simply assumed to be the same for the two DSSs but with opposite signs due to the opposite surface magnetic moments in the AFM TI. The $R_w$ term in $h_{t(b)}$ comes from the warping effect imposed by the rotation symmetry $C_{nz}$~\cite{Fu2009warping, Naselli2022Magnetic}.  As for the coupling term $h_{\mbox{\tiny{coup}}}$, it should be emphasized that besides the isotropic coupling up to $k^2$ order~\cite{Shan2010,Lu2010Massive,Sun2020prb,Lei2020pnas,Wang2023Dirac,wang2023three}, we have taken into account a symmetry-allowed anisotropic coupling, namely, the $R_a$ term. Remarkably, we have shown that the introduction of the anisotropic $R_a$ term can give birth to $2n$ Dirac-cone states located away from the $\Gamma$ point, as schematically shown in Fig. 1(c). These Dirac cones are termed intertwined Dirac cones, since they are induced from the hybridization of top and bottom DSSs. Furthermore, based on the intertwined Dirac-cone states, a high-Chern-number phase with $C=n$ can be achieved  by tuning the potential $U$~\cite{fan2023intertwined}, which paves the way for designing high Chern flat bands by twisting the AFM TI thin film as we show below.

\begin{figure}[htbp]
    \includegraphics[width=3.4in]{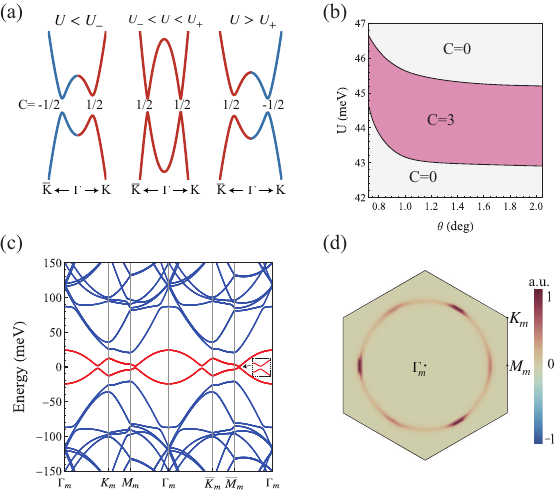}
    \caption{(a) Band structures with increasing electric potential $U$ for the two intertwined Dirac-cone states located along the $\overline{K}$-$\Gamma$-$K$ path in the original untwisted case, where the two Dirac states have the same (opposite) fractional Chern number for $U_-<U<U_+$ ($U<U_-$ or $U>U_+$). (b) Typical phase diagram with the twist angle $\theta$ and the electric potential $U$ for a twisted AFM TI thin film preserving the $C_{3z}$ symmetry. A high-Chern-number state with $C=3$ in the intermediate regime of $U$ persists for the concerned twist angles. Representative band structure (c) and corresponding Berry curvature distribution (d) in the moir\'e BZ for the high-Chern-number ($C=3$) state hosting two quasiflat bands near the Fermi level under a very small twist angle of $0.75^\circ$. In the above numerical calculations, the parameters are chosen typically as follows: lattice constant $a=2$ $ \mathrm{\AA}$, $m=0.04$ eV, $U=0.045$ eV, $v=1$ $\mathrm{eV\ \AA}$, $\Delta=0.02$ eV, $B=60$ $\mathrm{eV\ \AA^2}$, $R_a=200$ $\mathrm{eV\ \AA^3}$, $R_w=200$ $\mathrm{eV\ \AA^3}$, and for simplicity only a $k$-independent constant coupling with a strength of $\Delta/2$ is considered in $T_j$'s ($j=1, 2, ..., 6$).}
    \label{fig2}
\end{figure}

Considering the fact that the most well-studied (magnetic) TIs up to now are Bi$_2$Te$_3$ and MnBi$_2$Te$_4$ family materials respecting the threefold rotational symmetry $C_{3z}$, henceforth we choose the $n=3$ case in our paper, and the main results should remain valid for other cases of $n$. Correspondingly, the parameters in our effective model calculations of our paper are chosen in the same order of magnitude as those used in Bi$_2$Te$_3$ and MnBi$_2$Te$_4$ family materials~\cite{Fu2009warping, Lu2010Massive, Liu2010, Naselli2022Magnetic}. Figure 1(b) schematically shows the twisted Brillouin zones (BZs) of the top (orange lines) and bottom (blue lines) surface layers, and the moir\'e BZ (purple lines), where high-symmetry points are explicitly labeled. The moir\'e reciprocal vectors $\vec{b}_{im}$ ($i=1,2$) are given by the difference between the rotated reciprocal vectors of the bottom and top surfaces as $\vec{b}_{im}=\vec{b}_{i,t}-\vec{b}_{i,b}$. The length of $\vec{b}_{im}$ can be obtained as $|\vec{b}_{im}|=8\pi\sin(\theta/2)/(\sqrt{3} a_0)$, where $a_0$ is the lattice constant of the surface layer. The Hamiltonian after the twist can be written as 
\begin{equation}
H_{\theta}=\left[\begin{array}{cc}h_{t,-\theta/2}  +U\sigma_0& T \\T^\dagger & h_{b,\theta/2}- U\sigma_0\end{array}\right],
\end{equation}
where $h_{t(b),\mp\theta/2}=R_{\mp\theta/2}^{\dagger}h_{t(b)}R_{\mp\theta/2}$, with $R_{\mp\theta/2}=e^{\pm i\theta \sigma_z /4}$. $T$ represents the spatially periodic intersurface moir\'e hopping potential, and it suffices~\cite{Lian2020prl} to Fourier expand it to the lowest order as
\begin{equation}
T=T_{0}+\sum_{j=1}^{6}T_{j}e^{i\vec{g_j}\cdot \vec{r}}.
\end{equation}
Here, $g_j$'s are the six smallest moir\'e reciprocal vectors which can be generated from $\vec{b}_{1m}$ by sixfold rotations, as shown in Fig. 1(d). In the moir\'e reciprocal lattice, the above approximation amounts to considering the couplings within the first-shell reciprocal lattices spanned by $g_j$'s.

\section{High-Chern-number quasiflat bands} For later reference, we first briefly review the emergence of high-Chern-number state in the untwisted case, the details of which can be found in Ref. \cite{fan2023intertwined}. As shown in Fig. 2(a), when the potential $U$ is smaller than $U_-$ or larger than $U_+$,  with $U_{\pm}$ given by \cite{fan2023intertwined}
\begin{equation}
U_{\pm}=\sqrt{v^2\Delta/B+[m\pm R_w(\Delta/B)^{3/2}]^2},
\end{equation}
the two intertwined Dirac-cone states (the other four Dirac-cone states are related by $C_{3z}$ rotations) located along $\Gamma\rightarrow K$ and $\Gamma\rightarrow \overline{K}$ directions, respectively, have opposite fractional Chern numbers $\pm 1/2$. As a result,  the total Chern number $C$ equals zero at the Fermi level, whereas when $U_-<U<U_+$, the above two Dirac-cone states have identical fractional Chern numbers ($C=1/2$), leading to a high-Chern-number state with $C=3$.

After the twist between the top and bottom surface layers, the original band structures will get significantly modified in the moir\'e BZ. First, the critical electric fields $U_{\pm}$ where topological phase transitions happen accompanying the gap closing-and-reopening processes of the intertwined Dirac-cone states are no longer fixed and instead change with $\theta$. A typical topological phase diagram as a function of $\theta$ and $U$ is presented in Fig. 2(b), where both $U_{+}$ and $U_{-}$ are found to increase with decreasing $\theta$. Second, since the moir\'e BZ is much smaller than the original BZ, the moir\'e band structures are expected to have a much reduced bandwidth with less dispersive bands, as will be discussed in detail below. Most importantly, for a very small twist angle ($\theta<1^\circ$), a $C=3$ high-Chern-number state hosting almost flat bands near the Fermi level can emerge. This is exemplified by the band structure of $\theta=0.75^\circ$ and $U=45$ meV, shown in Fig. 2(c), where the bandwidths of the highest valence band (HVB) and the lowest conduction band (LCB) become smaller than 25 meV. We have also plotted the corresponding Berry curvature distribution in the moir\'e BZ in Fig. 2(d), and it can be seen that Berry curvatures are still mainly concentrated around the intertwined Dirac points located along the $M_m$-$\Gamma_m$-$\overline{M}_m$ directions.

\begin{figure}[htbp]
    \includegraphics[width=3.2in]{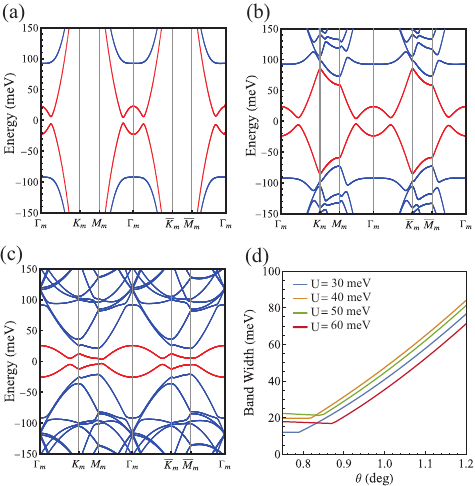}
    \caption{Band structures in the moir\'e BZ of the twisted AFM TI thin film with gradually reduced twist angels of  (a) $\theta=2^\circ$, (b)$\theta=1.2^\circ$, and (c) $\theta=0.75^\circ$. A significant reduction of the dispersion of the highest valence and lowest conduction bands near the Fermi level can be clearly seen, which become nearly flat in (c). (d) The evolution of the bandwidth of the highest valence band with decreasing twist angle $\theta$. Besides $\theta$, the other parameters in the numerical calculations are the same as those in Fig. 2.}
    \label{fig3}
\end{figure}

\section{Twist-engineered band structures} By tuning the twist angle, the band structure of the AFM TI can be engineered in two aspects. On the one hand, the bandwidth can be reduced by decreasing $\theta$, and the bands could become nearly flat for quite small $\theta$.  Figures 3(a)$-$3(c) show the evolution of a typical band structure for $U=50$ meV with gradually decreased values of $\theta=2^\circ$ [Fig. 3(a)], $\theta=1.2^\circ$ [Fig. 3(b)], and $\theta=0.75^\circ$ [Fig. 3(c)], where a significant reduction of the band width can be clearly seen from the particle-hole symmetric HVB and LCB. Moreover, we have explicitly plotted the band width of the HVB with decreasing $\theta$ for different values of the potential, as shown in Fig. 3(d). All of them exhibit a monotonous decrease of the band width with gradually reducing $\theta$ until it reaches a small critical value, beyond which the band becomes quite flat with its width around 20 meV.  On the other hand, according to the phase diagram in Fig. 2(b), topological phase transitions between trivial ($C=0$) and nontrivial ($C=3$) could be induced by simply tuning the twist angle.

Further, we show that the topological property of the quasiflat bands formed by the HVB and LCB near the Fermi level at small twist angles can be easily tuned by the electric field. With increasing the potential $U$, two successive gap closing-and-reopening processes are found to occur at $U_-$ and $U_+$ from the intertwined Dirac states located along the $\Gamma$-$\overline{M}_m$ (and other two $C_{3z}$-symmetry related paths) and $\Gamma$-$M_m$ directions, respectively, in the moir\'e BZ. As an example, we have plotted the band structures at $\theta=0.8^\circ$ for three representative values of $U$, namely, $U<U_-$ [Fig. 4(a)], $U_-<U<U_+$ [Fig. 4(c)], and $U>U_+$ [Fig. 4(e)]. The corresponding Berry curvatures of the HVB are presented in Figs. 4(b), 4(d), and 4(f), respectively. The Berry curvature around the intertwined Dirac state along $\Gamma$-$\overline{M}_m$ reverses its sign from negative to positive across the transition point of $U_-$, as can be seen from Figs. 4(b) and 4(d). This contributes a total change of $+3$ for the Chern number of the occupied bands, thus driving a phase transition from a trivial flat-band state with $C=0$ to a high-Chern-number $(C=3)$ nontrivial flat-band state. Similarly, the sign change  of the Berry curvatures from positive to negative around the intertwined Dirac states along $\Gamma$-$M_m$ across $U_+$ changes the Chern number by $-3$, and the flat-band system returns to the  trivial $C=0$ state.

\begin{figure}[htbp]
    \includegraphics[width=3.2in]{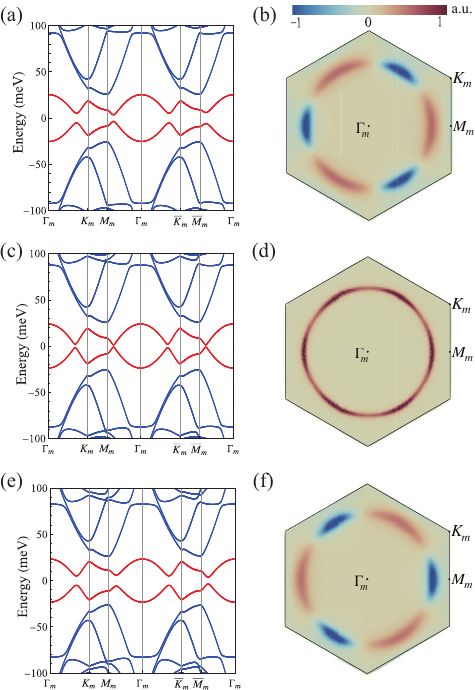}
    \caption{Electrically tunable flat-band structures (first column) and corresponding Berry curvatures (second column) of the twisted AFM TI thin film at $\theta=0.8^\circ$ with $U=40$ meV ($C=0$, first row), 45 meV ($C=3$, second row), and  50 meV ($C=0$, third row), respectively. The other parameters in the numerical calculations are the same as those in Fig. 2.}
    \label{fig4}
\end{figure}

\section{Summary and discussion} In summary, based on the effective model analysis, we have proposed realizing high-Chern-number ($C=n$) flat-band states near the Fermi level in twisted AFM TI thin films preserving an  $n$-fold rotational symmetry. The intertwined Dirac-cone states induced by anisotropic inter-surface coupling are found to play a crucial role in forming these nontrivial flat bands through the twisting procedure. Furthermore, we have also demonstrated that an out-of-plane electric field could drive topological phase transitions between nontrivial ($C=n$) and trivial ($C=0$) flat-band states. Our work not only sheds light on the significance of the recently proposed intertwined Dirac-cone states, but also open a new avenue to realize high-Chern-number flat bands.

It is noteworthy that the $C=3$ phase has already been confirmed in MnBi$_2$Te$_4$$/$(Bi$_2$Te$_3$)$_{m}/$MnBi$_2$Te$_4$ ($m=0$, $1$, $2$) heterostructures through first-principles calculations~\cite{fan2023intertwined}, and thus the proposed flat-band states could hopefully be realized in these materials by moir\'e engineering. As for the experimental realization, the electric field (displacement field) can be applied and changed through the commonly used dual-gate technique, where both the  displacement field and charge density can be simultaneously tuned~\cite{chen2020tunable, cao2020tunable}. To be more specific, the electric field potential for the typical high-Chern-number flat-band state of our work lies in the range between 40 meV and 50 meV, and when taking a 4-septuple-layer MnBi$_2$Te$_4$ thin film with a thickness of $\sim$5 nm ($\sim$1.36 nm per septuple layer) as an example, the required electric field corresponds to the displacement field between 0.08 V/nm and 0.1 V/nm (a typical dielectric constant and screening factor of $\sim$ 10 has been considered), which is easily accessible in experiments \cite{chen2020tunable,cao2020tunable}. Notably, when considering lattice relaxation effects~\cite{Nam2017, Leconte2022}, such as an enhancement of the Fermi velocity and the change of interlayer distance with modified interlayer couplings, the corresponding critical electric field strength may get slightly changed. Nevertheless, since the Chern-insulator state corresponds to a charge neutrality single-particle gap, no correlation effects or fine-tuning fractional fillings are required, thus ensuring its stability and feasibility for observation. Interestingly, if electron-electron interactions are considered, exotic states such as a fractional Chern insulator and chiral superconductivity may emerge in these systems, which will be left for future work.\\\\

\begin{acknowledgements}
This work is supported by National Key Projects for Research and Development of China (Grants No. 2021YFA1400400 and No. 2023YFA1407001), the Fundamental Research Funds for the Central Universities (Grant No. 020414380185), Natural Science Foundation of Jiangsu Province (No. BK20200007), the Natural Science Foundation of China (No. 12074181, No. 12104217, and No. 92365203), and the Department of Science and Technology of Jiangsu Province (BK20220032) \\
H.W and Y.F contributed equally to this work.
\end{acknowledgements}

\bibliography{ref}

\end{document}